\documentclass[EN, showkeys,showpacs]{revtex4}
\usepackage{graphicx}
\usepackage[utf8]{inputenc}
\usepackage{wrapfig}
\usepackage{algpseudocode}
\usepackage{epsfig,color}
\usepackage{latexsym,amsmath,amssymb}
\usepackage{amsthm}

\def\deltap{\delta^+}
\def\Dij{\Delta_{ij}}
\def\deltap{\partial^+}
\def\deltapp{\partial^{++}}

\def\deltal{\Delta^+}

\def\sit{s_i^t}
\def\yit{v_i^t}
\def\y{v}

\begin{document}
\title{Contamination source inference in water distribution networks} 

\author{E. Ortega}
\affiliation{Facultad de F\'{\i}sica, Universidad de la Habana.}
\author{A. Braunstein}
\affiliation{DISAT, Politecnico di Torino, Corso duca degli Abruzzi 24, Turin, Italy
HUGEF, Via Nizza 52, Turin, Italy, Collegio Carlo Alberto, Moncalieri, Italy.}
\author{A. Lage-Castellanos}
\email{ale.lage@gmail.com}
\affiliation{Facultad de F\'{\i}sica, Universidad de la Habana.}



%
\begin{abstract}
 We study the inference of the origin and the pattern of contamination in water distribution networks.
We assume a simplified model for the dyanmics of the contamination spread inside a water distribution network, and assume that at some random location a 
sensor detects the presence of contaminants. We transform the source location problem into an optimization problem by considering discrete times and a 
binary contaminated/not contaminated state for the nodes of the network. The resulting problem is solved by Mixed Integer Linear Programming. 
We test our results on random networks as well as in the Modena city network.
\end{abstract}





\maketitle

\section{Introduction}

Fluid networks are ubiquitous and fundamental. They come in a variety of sizes and shapes, from circulatory and lymphatic systems in vertebrates, to ocean conveyor belts and sea currents, passing by industrial fluid networks, water distribution systems in cities and hydrographic basins. In all of these cases, the spread of contamination, defined as the presence in the fluid of non desired material, is problematic. We study how to infer the position in the network at which a contaminant first appeared after the observation of its presence in some distant nodes of the network. This is a relevant question at least in industries and cities, where pipe networks are not susceptible of being explored or sensed at every node, and where future or further damage can be prevented from the knowledge of the contamination source.

The study of contamination origin has received a lot of attention in the near past. Several approaches have been developed \cite{Ale,luo2013identifying,milling2012identifying,spinelli2013source,zhu2016information,luo2014identify,lokhov2014inferring} with the goal of infering the patient zero after observing an epidemic outbreak. In \cite{jiang2017identifying} there is a state of the art of studies made in related topics. However, epidemic models in networks differ from fluid networks in that they are essentially stochastic. Pipe systems can have some source of stochasticity, like turbulence, but to the scope of the present study the movement of the fluid is considered deterministic. The simplicity of fluid networks is compensated by the fact that we will try to infer the origin with very little information, namely the observation of contamination in one or two nodes of the system.

There is a variety of questions related to the inference of contamination in water distribution systems, and a variety of methods and approaches. Some researchers \cite{But} have focused on the inference of the pattern at the origin, after observation of a contamination pattern at an observation node, but under the assumption that both the origin and the observation points are known. In such cases, the target is to reproduce the real-valued time dependent concentration of contaminants. Instead, we will pay more attention to discovering the unknown source of contamination. 

Within the scope of detecting the source of contamination, many researchers make use of extensive forward in time simulations (usually resorting to EPANET) \cite{preis2006contamination,perelman2010bayesian,cristo2008pollution,huang2009data,hu2015mapreduce,guan,tao2012identification} to test different possible origins, and compare their predicted patterns with the observed one, or to create a rule that allows them to do inference later on. Some consider that there is a source of stochasticity coming from the uncertainties in the demands on the network at any given time \cite{huang2009data,wang2011bayesian} or on the reliability of the sensors \cite{propato}, and try to infer the contamination source in such a context, sometimes resorting to Monte Carlo simulations \cite{wang2011bayesian} or Bayesian methods \cite{huang2009data,wang2011bayesian,propato}. Some have attempted the detection of the origin by directly reversing the flow dynamics in the network \cite{REVERSE}.

Our approach has some contact points with some of these. As in \cite{huang2009data,wang2011bayesian,propato} we formalize our method in Bayesian terms, although we don't consider stochasticity in the demand or on the sensors. We assume a simplified version of the problem, with known deterministic dynamics, where time is discretized and contamination is reduced to be binary (contaminated or not) as in \cite{huang2009data}. Our inference method will rely on a neat mathematical approach using linear programming to find the most effective explanation for a contamination event. The linear programming optimization will arise from an Occam's Razor-like argument by the assumption that contamination is a rare event. 

Section 1\ref{sec:Assumptions} establishes the simplifications assumed throughout this work in the modeling of the water distribution networks, and the treatment of their dynamics. A precise statement of our mathematical problem appears in section 2\ref{sec:optimization}, where it is also transformed into an optimization problem. The results obtained in random models of cities appear in section 3\ref{sec:results}, and everything is summarized at the end.

\section{1. Assumptions and simplifications}
\label{sec:Assumptions}
Fluid networks can be grouped in two big classes, non-cyclic and cyclic regarding on whether any piece of fluid is allowed to pass by a given position in the network more than once. Circulatory systems and the ocean conveyor belt are examples of cyclic networks. On the other hand, many industrial and city networks are non-cyclic. The procedure presented by us will focus on non-cyclic networks. Extensions to cyclic networks might be simple, but have not been explored so far.

Furthermore, we will focus on water distribution networks (WDN), this is, the system of pipes, pumps, elevated tanks, junctures and consuming points that characterize a standard city clean water network. The behavior of such networks are described by the Todini-Pilati equations \cite{Todini}, in which two fundamental constraints, the conservation of mass and the conservation of energy, are grouped in matricial form in the following way
\begin{equation}
\left[\begin{array}{ccc}
\ A_{pp}&\ A_{pn}\\
\ A_{np}&\ 0\\
\end{array} \right]\left[\begin{array}{ccc}
\ Q\\
\ H\\
\end{array} \right] = \left[\begin{array}{ccc}
\ -A_{p0}H_0\\
\ q\\
\end{array} \right]. \label{eq:todini}
\end{equation}
On right-hand side, the flow of demand $q$ at the consumption nodes is given, as well as the head pressure $H_0$ at the pumps and elevated tanks of the city. On the left-hand side, the flow in each pipe $Q$ and the pressure in internal junctures $H$ are to be found in order to satisfy the energy and continuity equations. Detailed explanations of each term and a discussion on the meaning and solution of the equations can be found in \cite{Gi09}. Despite their apparent simplicity, these equations are non-linear, since the elements of the matrix $A_{pp}(i,i)=R_i|Q_i|^{n-1}$ depend on the flows $Q$ we are looking for.

Inside the matrices  $A_{pp}$, $A_{pn}$, $A_{np}$ and $A_{p0}$ resides the information about the topology of the network, the radii of the pipes as well as the drag coefficients inside the pipes. In principle, all these parameters are known (at least approximately) and the flows $Q$ and the fluid velocities $v$ inside the pipes, can be found using free and public modeling software like EPANET \cite{epanet}. This software allows also the study of the diffusion of contaminants in the network.

\begin{figure}
\includegraphics[width=5cm,height=5.5cm]{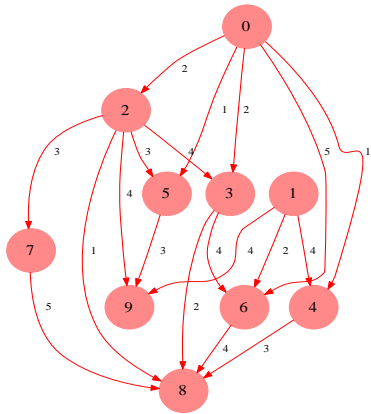}
\caption{\label{fig:redjuguete} Schematic representation of an abstract water distribution network. Numbers inside nodes are labels. Numbers besides edges are time delays $\Dij$.}
\end{figure}

However, we will concentrate our effort on studying the direct  (forward in time) problem and inverse (inference)  problem in a simplified version of the water dynamics. In order to do this we will assume the following simplifications:
\begin{itemize}
 \item {\bf discrete time:} once the fluid velocities $v$ are known, we obtain the pipe time $\Delta = \mbox{Round} ( \frac{L}{v})$ from the pipe lengths $L$. We will consider this time $\Delta$ an integer, using a suitable discretization. This discretization is used for describing the states $\sit$ of each site $i$, at time $t$, in the network.
\item {\bf binary contamination:} at each time the state of node $i$ can be either clean, $\sit=0$, or contaminated, $\sit=1$. We shall make the distinction between the state $\sit\in\{0,1\}$ of a node, and the variable $\yit\in\{0,1\}$ signaling whether node $n_i$ is being actively contaminated by an external source in time $t$. In other words  $\yit=1$ marks node $n_i$ at time $t$ as origin of the contamination. 
\item {\bf deterministic: } we will suppose all real parameters of the system (head pressures, drag coefficients, pipe lengths, demand flow, etc.) as known in detail and fixed in time, such that the system operates in a stationary regime. Stationarity was assumed for simplicity, but is not essential to our method, and can be readily lifted.
\item {\bf WDN is a graph:} any water distribution network is representable as a directed, non-cyclic graph with timed edges  $G(V,E,\Delta)$:
 \begin{itemize}
  \item $V=\{n_i | i\in[0,1,\ldots N]\}$ is the set of all nodes (vertexes) in the graph, 
  \item $E=\{(n_i,n_j)\}$ is the set of directed $n_i \to n_j$ edges with time delay $\Dij$
  \item $\Dij$ is the time delay of each pipe. When $(n_i,n_j) \in E$, $\Dij>0$ and $\Delta_{ji} = - \Dij$. If  $(n_j,n_i) \notin E$ then $\Dij = \infty$.
\end{itemize}
\end{itemize}

In practice the graph $G(V,E,\Delta)$ is constructed using the topological network data (which nodes are joined by pipes). But this is not enough. We also need the stationary solutions of the Todini-Pilati equations, since we need the time $\Dij$ the fluid takes to travel through pipe $(n_i,n_j)$ that depends on the velocities of the fluid. In this sense, $G(V,E,\Dij)$  summarizes topological as well as dynamical information of our system.

\subsection{Simplified dynamics for direct problem}
The forward in time evolution of contaminants in the network is ruled by the following consistency equation between the state of node $n_i$ and the sates of it's predecessors $\deltap j=\{ n_j | (n_j,n_i) \in E \}$:
\begin{eqnarray}
\sit &=&m(s_{\deltap i}^{t-\Delta} ,\yit) \equiv \yit \bigvee_{j\in \deltap i} s_j^{t-\Delta_{ji}}  \label{eq:sit}
\end{eqnarray}
The meaning of this equation is quite obvious. The  node $n_i$ at time $t$ can be contaminated ($\sit=1$) if either it is the origin of the contamination ($\yit =1$) or it is receiving contaminants from nodes connected to it $\exists_{j\in\deltap i} s_j^{t-\Delta_{ji}} =1 $.

\subsection{Graph-time expansion}

\begin{figure}
\includegraphics[width=7cm,height=5.5cm]{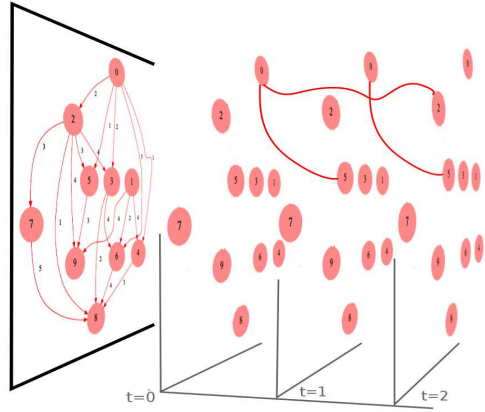}
\caption{\label{fig:graphexpansion} Representation of the time-extended graph derived from the graph in figure \ref{fig:redjuguete}.}
\end{figure}

Equation  (\ref{eq:sit}) suggests that the dynamics of contamination can be described in a time-extended graph. In it nodes $s_i^t$ are not only defined by their spatial location $i$ in the WDN but also by a (discrete) time index $t$. We define directed edges to be existent between nodes $(s_j^{t-\Delta_{ji}} \to s_i^t)$ if a pipe connects node $j$ with node $i$ with a time delay of $\Delta_{ji}$. In figure \ref{fig:graphexpansion} we show the first three time slices of the network of figure \ref{fig:redjuguete}. To avoid overloading the figure, we only represented three edges, two of them connecting physical node $0$ to physical node $5$, in a $\Delta=1$ time delay, and one connecting node $0$ to node $2$ in $\Delta = 2$ time steps. 

The time-extended graph will be the starting point of our construction, allowing us to simulate simplified dynamics through eq. (\ref{eq:sit}). To test our inference procedure, we generate random graphs with the desired properties (directed, non-cyclic and timed edges) pick up a random source node and a random pattern $\yit = (0,0,0,1,1,0,1,1,0,0,\ldots)$, and use equation (\ref{eq:sit}) to track the evolution throughout the network. We will also pick, among the contaminated nodes, one or two that will be used as sensor, and whose evolution in time $s_o^t = (0,0,0,0,0,0,1,1,0,1,1,\ldots)$ will be accessible to the inference procedure that we explain next.

\section{2. Inference as optimization problem}
\label{sec:optimization}
The inverse problem we want to solve can be stated as follows. Given a water distribution network, characterized by a 
graph $G(V,E,\Delta)$, and a set $O= \{o_1,o_2,\ldots\}$ of observed nodes with their patterns $s_{o_i}^t = (0,0,0,1,1,\ldots)$, 
find the most probable set $P=\{p_1,p_2,\ldots\}$ of nodes and their patterns $\y_p^t = (0,0,1,1,\ldots)$ for the origin 
of the contaminants in the network. When convenient we will denote both the observed nodes and the contamination patterns 
by $O$ and $P$ in each case.

In the estimation theory language, and using the Bayes formula, we want to maximize, over the set of origins $P$, the probability 
\[Prob(P|O)   = \frac{Prob(O|P) Prob(P)} {Prob(O)}.\]
To the scope of maximization, the denominator $P(O)$ is unknown but irrelevant. On the other hand, the probability of observing contamination $O$ given a set of sources $P$ 
\[ Prob(O|P) \:\: =\:\: \sum_{\sit\notin O} \prod_i \prod_t \delta_{\sit,m(s_{\deltap i}^{t-\Delta} ,\yit)}
\]
is either one or zero, depending on whether or not the deterministic evolution of contamination in time following equation (\ref{eq:sit}) reproduces the observed pattern. This is a consequence of the deterministic assumption made for the dynamics of the network.

Therefore, the maximum of $Prob(P|O) \propto Prob(O|P) Prob(P)$ is found among the origins that are consistent with the observation, that we will symbolize as $P\to O$. In other words, we will estimate the origin of the contamination as the set of nodes and patterns satisfying
\[ \hat P = \underset{P:P \to O}{\mbox{argmax}}\:\: Prob(P).
\]
Now we need to fix a prior for the patterns. Assuming that the contamination is a rare event, will be considered as most probable the explanations requiring the smallest amount of nodes and time laps involved in the original contamination, therefore, the smallest set $P$.

Using equation (\ref{eq:sit}) recursively, we can write the consistency equations between the observation at any node, and the original contamination variables $\yit$ at nodes above it in the graph,
\begin{eqnarray}
\sit =& M(v_{\deltapp i}^{t-\Delta_c}, \yit) &   \equiv \yit \bigvee_{j\in \deltapp i}\bigvee_{c\in C_{ij}} v_j^{t-\deltal_c } \label{eq:sit_caminos}.
\end{eqnarray}
The reason why this recursive substitution of eq. (\ref{eq:sit}) in itself has an end point is that we are considering non-cyclic networks, so the observation at node $s_o^t$ can only come from a finite number of nodes and times in the graph. In expression (\ref{eq:sit_caminos}), $\deltal_c$ stands for the time delay of path $c$ from node $n_j$ to node $n_i$, while $\deltapp i$ stands for all nodes $n_j$ that can access node $n_i$ by some path $c$ in graph $G(V,E,\Delta)$. The expression $v_{\deltapp i}^{t-\Delta_c}$ inside the definition of function $M(\cdot,\cdot)$ represents all element $v_j^{t-\deltal_c }$ on the right-hand side of the equation.

The condition $P\to O$, read as ``{\it origins $P=\{p_1,p_2,\ldots\}$ causes observation $O$}'' is equivalent to the condition
\[\forall_{o\in O} \; \forall_t \:\: s_o^t = M(v_{\deltapp o}^{t-\Delta_c}, v_o^t).\]

\subsection{Graph reduction} As we want the smallest set $P$, any node that is not present on the right-hand side of this equation, will be assumed not to be an origin $\yit = 0$, since it is irrelevant to explain the observation. Furthermore, the nodes that do appear on the right-hand side of one of these  equations, but whose left-hand side is a clean observation $s_o^t=0$, are forced by this equation to also be clean, $v_j^{t-\deltal_c }=0$. Therefore, in order to explain the observation we remain with the observed nodes variables $v_o^t$ that are seen as contaminated $s_o^t=1$ and those $v_j^{t-\deltal_c }$ connected to them by equation (\ref{eq:sit_caminos})  that are not connected to non contaminated observations. All this reduction of the valid nodes is equivalent to what is done in the contamination source pruning of \cite{propato}.

\begin{figure}
\includegraphics[width=6cm,height=6.5cm,angle=270]{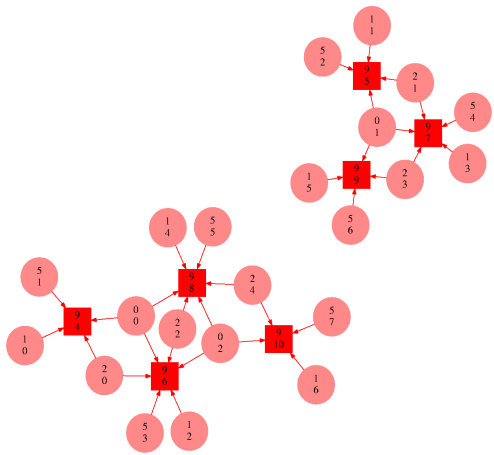}
\caption{\label{fig:reducedgraph} Each node is labeled by two numbers, the upper one is the original node number in the graph (figure \ref{fig:redjuguete}) while the lower one stands for the time bin of the variable $\yit$ it represents. Square nodes correspond to observed state variables $v_o^t$, while circular nodes correspond to the remaining variables (see text).}
\end{figure}

We can represent the relation between the remaining variables as a graph, in which each contaminated observed node variable $v_o^t$ is connected to the non-observed nodes that remain as possible explanations of the observed state, as in figure \ref{fig:reducedgraph}. Furthermore, re-labeling the variables in this representation as $y_k$ for the observed (square) nodes $v_o^t$ and $x_i$ for the non-observed (circular) nodes $v_j^{t-\deltal_c }$, the constraints imposed by eq. (\ref{eq:sit_caminos}) can be stated as
\begin{equation}
\forall_{y_i\in Y} \quad  y_i + \sum_{x_k\in \partial {y_i}}  x_k  \geq 1 \label{eq:rest}
\end{equation}
forcing each observed node itself, or one of its neighbors to be the origin of the contamination. The condition $\hat P = \underset{P:P \to O}{\mbox{argmax}}\:\: Prob(P)$ now can be rewritten as
\begin{equation}
(\hat x ,\hat y) = \underset{x,y}{\mbox{argmin}} \left(\sum_{i} x_i + \sum_{j} y_j \right) \label{eq:min}
\end{equation}
under the constraints given by eq. (\ref{eq:rest}).

Both the objective function and the constraints are linear, and therefore susceptible of being solved by integer/linear programming methods(LP). Since coefficients are integers, more than one optimal solution might exist. The set of all solutions can be found by recursively slightly altering the coefficients of the variables in the objective function (see \cite{THESIS}). Once a solution is known, we get a set of variables $y$ and $x$ with value 1, while the rest is zero. Mapping back $y_k$ and $x_i$ to the original variables $v_o^t$ and $\yit$, we claim the solution to be the most likely pattern of contamination. 

At this point it is important to clarify that the optimization made by LP is only in time, meaning that any two solution with the same number of active times for the contamination are equivalent. In other words, the optimization does not distinguish whether the source nodes of the time-extended graph correspond to the same node in the water network, or to different nodes. This is relevant, since the basic assumption made, namely that contamination is a rare event, would also imply that solutions with one real node of the water network might be preferable to those involving more nodes, even in the cases where the latter involves less time-expanded nodes. The current procedure can not solve this issue neatly. We partially circumvent it by choosing among the solutions those involving less real nodes of the water distribution network.

 
We have shown how to transform the inference of contamination into a treatable optimization problem. Let us outline the whole procedure here:
\begin{enumerate}
 \item {\bf Input: } Start from the given WDN and the set of pipe times obtained from the solution of the Todini-Pilati equations for your system, and some observed contamination pattern. (For this step one probably will use EPANET or other similar software, as we did for the network of Modena city. In the case of random networks, this step was not required, since the time intervals were directly generated from a probability distributions.)
 \item {\bf Graph-time expansion: } Using the speeds along the pipes, and the distances of the pipes, compute the time delays $\Delta_{ji}$. Use a suitable discretization of the times to create a time-extended graph as explained in the previous section. (We programmed this step in {\it python}.)
 \item {\bf Graph reduction: } Starting from the time-extended graph, create a new graph by connecting every node $s_i^t$ to all other nodes $s_k^{t_k}$ such that a path \[ (s_k^{t_k} \to s_{k1}^{t_{k1}}),(s_{k1}^{t_{k1}} \to s_{k2}^{t_{k2}}),\dots,(s_{kn}^{t_{kn}} \to s_{i}^{t}) \] exists. Reduce this graph (as in figure \ref{fig:reducedgraph}) by retaining only the observed contaminated nodes and those connected to them that could possibly explain the contamination. (We programmed this step in {\it python}.)
 \item  {\bf Optimization: }Use linear programming to find the most efficient explanation for the contamination observed. (We used {\it lpsolve} solver in linux.)
\end{enumerate}

Next we discuss two typical situations where our method fails and succeeds. In the section after, we test the efficiency in three different situations: many unrealistic random topologies water distribution networks, a real city (Modena in Italy) and finally many realistic random cities statistically similar to Modena.  

\section{Simple small examples}
\label{sec:examples}

Let us consider two separate examples, using the toy network of figure \ref{fig:redjuguete}. The first example is the observation of a contamination  at sensed node $7$ (bottom left, painted in {\color{blue} blue} color), with pattern $s_7 = (0,0,0,0,0,1,0,0,1,0,0)$. Given that there is only one path in the network arriving at node 7, it is obvious that contamination should have happened along that path, this is, either at 7 itself, at node 2, at node 0, or at a combination of them. Our algorithm (and logic) will produce the following solutions (only representing the non-zero modes):
\begin{eqnarray*}
 S &=& \{ \color{blue} (v_7^5, v_7^8), (v_2^2, v_2^5), (v_0^0, v_0^3) \\
 && (v_7^5, v_2^5), (v_7^5, v_1^3) \\
 && (v_2^2, v_7^8), (v_2^2, v_1^3) \\
  && (v_1^0, v_2^5), (v_1^0, v_7^8)  \}
\end{eqnarray*}

While all those solutions involve two nodes of the time-extended graph and are equivalent to our optimization code, only the first three involve one real node of the WDN. We call this solutions 1-node-solutions. Assuming that contamination is a rare event, we will select the subset of 1-node-solutions, whenever it is not empty, as the most probable contamination source. In this case, however, neither the algorithm nor logic can reduce our uncertainty among which of the three nodes $(7,2,0)$ was the most probable origin. This is a difficulty that arise also in many other contamination inference algorithms \cite{preis2006contamination,huang2009data,tao2012identification,REVERSE}.

\begin{figure}
\includegraphics[width=4.5cm,height=4.0cm]{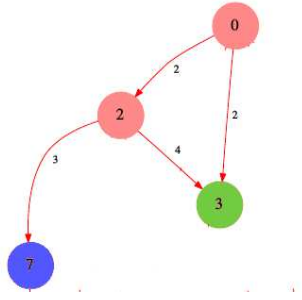}
\caption{\label{fig:smallexample} Part of the Graph from figure \ref{fig:redjuguete}. Blue (node 7, bottom left) and green (node 3, bottom right) nodes are nodes where two different contamination events are sensed.}
\end{figure}
The second example is the observation of a contamination at sensed node $3$ (bottom right, painted in {\color{green} green} color), with pattern $s_3 = (0,0,0,0,0,1,0,0,0,1,0)$. Now there are  two paths in the network arriving at node 3. As before, we can explain the contamination with single node solutions as
\begin{eqnarray*}
 S &=& \{ {\color{green} (v_0^3)}, (v_2^1, v_2^5), (v_3^5, v_3^9) \}
\end{eqnarray*}
The outcome of our algorithm, however, will be only $(v_0^3)$, since it minimizes the number of nodes in the time extended graph required to explain the contamination effect. This illustrates both, the power and the risk of our algorithm. We expect our optimization procedure to return non-trivial inference whenever there is more than one path connecting the real contamination source with the observation nodes. At the same time, complex in time patterns (with contamination lasting for more than one time interval) are susceptible of being over-simplified. That would be the case of the previous example, if the real contamination origin had been node 2, a situation that our algorithm would disregard as sub-optimal.

We could study mathematically the probability of such situations appearing in water distribution networks, but we leave this for a future study on the optimal sensor distribution. For the optimality of sensor placement, exploiting multi-path will be crucial. We next characterize ``experimentally'' our method by trying it on many artificial networks.

\section{3. Random distribution networks}
\label{sec:results}

In the following we will treat only contaminations events that take place in a single node of the network. Therefore if the true contamination pattern is a solution to of our optimization problem, then it is in the 1-node-solutions. Despite already being a reduction, the 1-node-solutions can contain many different explanations for the contamination. For instance if we have a WDN that is simply a single chain of nodes with water moving along, and a sensor at the end of it, the method will find all nodes in this line are equally good 1-node explanations of the observed pattern.

We define the efficiency of our method in two different ways. First, by the probability of finding the real contamination pattern in the 1-node-solutions. Second, by the probability that the real contamination pattern is one randomly selected 1-node solution. This second measure is more demanding, since in order to be efficient, the method should find few optimal solutions, while the first measure is insensitive to the multiplicity of solutions.

\begin{figure}
\includegraphics[width=0.35\linewidth,angle=270]{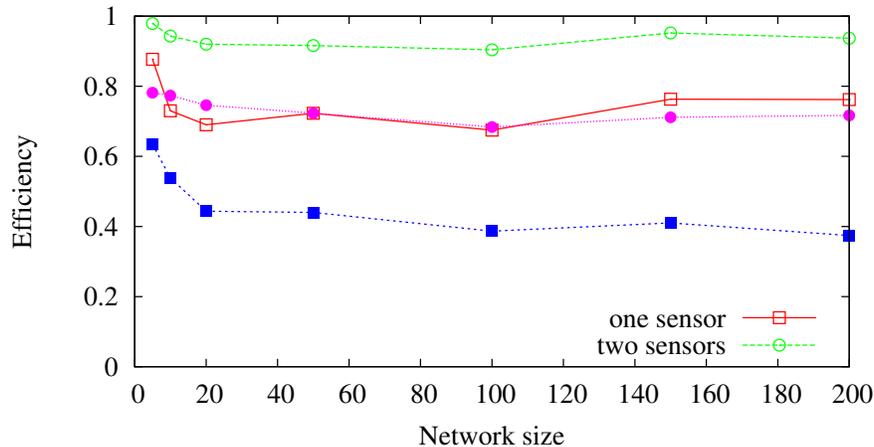}
\caption{\label{fig:network_bi_1} Efficiency of the inference method with one and two sensors in a system with a compact contamination pattern. Squares correspond to one sensor cases, while circles correspond to two sensors. Open points correspond to the first efficiency measure, while the full points correspond to the second (see first paragraph in this section). Each data point is an average of 1000 different contamination events in different systems.}
\end{figure}

To explore the efficiency  of the method we first test it on random graphs with WDN characteristics. These random graphs (as the one in figure \ref{fig:redjuguete}) are Erdos-Renyi graphs \cite{Erd}  in which a directionality and a time delay are enforced in each pipe.
Figure \ref{fig:network_bi_1} shows the efficiency of the method versus the number of nodes in the system. The origin of the contamination is taken to be a single randomly selected node $i$, and we spread from it a compact contamination pattern (contamination in consecutive times steps, lasting a randomly selected time between 2 and 10) at $t=0$ $v_i=(1,1,1,0,\ldots)$. The selection of sensors is made in the following way: the sensed node is picked randomly from the set of nodes touched by the contamination but guaranteeing that it is not connected with previously chosen sensors (because in this case the contamination in one of them can explain the contamination in the other). If this condition can not be satisfied, then the second sensor is picked randomly from the nodes through which contamination does not passes (the information of a node not being contaminated is also informative, since it allows for the removal of all its neighbors from the set of possible explanations).   

We could have done simulations for bigger graphs, but the trend seems already to be stable in the range $20-200$. Let us underline that 200 nodes is already a size comparable to that of small cities, or at least of some industries. However, we didn't push forward in this direction, since an Erdos-Renyi random graph is hardly a credible topology for real water distribution networks (see Modena and Modena-like random cities in the next sections).

The efficiency practically remains constant for 50 or more nodes. With two sensors in the system the method is nearly perfect since practically in every case of study the real contamination pattern was obtained among the optimal solutions found. However, in both cases, the inference method finds many 1-node solutions for the optimization problem, reducing the second efficiency measure. 

\begin{figure}
\includegraphics[width=0.35\linewidth,angle=270]{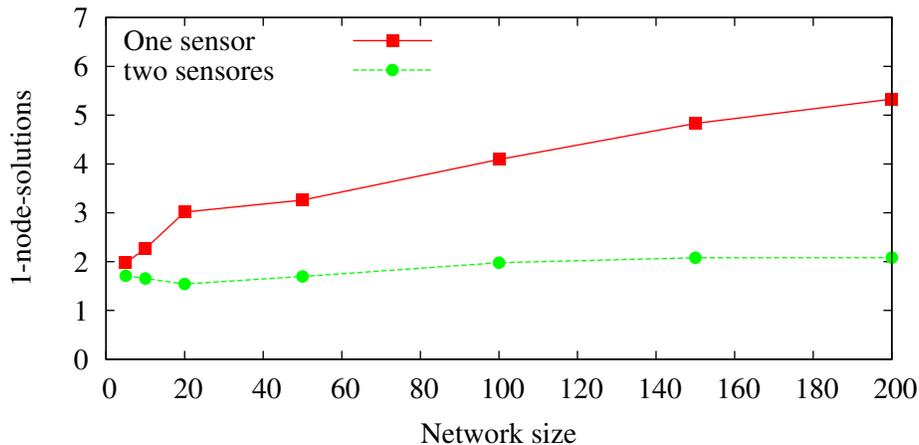}
\caption{\label{fig:network_bi_2} Comparison between the number of solutions which involve only one node of the graph $G$ with one or two sensors in the system. Each data point is an average of 1000 different contamination events on different systems.}
\end{figure}

In any case the number of solutions found by the method is also a relevant parameter to characterize its efficiency. As we mentioned at the end of section 2\ref{sec:optimization} the method can not determine the solutions which involve less number of nodes the real water distribution network (graph $G$) directly from optimization (\ref{eq:min}), but rather those involving less nodes of the time-extended graph. Given a set of optimal solutions we can further reduce this set by selecting among them those involving fewer real nodes in the network. In figure \ref{fig:network_bi_2} we show the number of optimal solutions which involve only one real node of graph $G$.


Observe that the number of 1-node-solutions is never higher than ten, this means that with one sensor in a network of 200 nodes the number of nodes which are possible source of the contamination are around 10, which is significantly smaller than the network size, but still not a reliable estimation method. However, with two sensors the average number of optimal solutions is between 1 and 2, which is quite a reduction of the uncertainty.  
\section{Modena city}

To get more realistic, we studied a real city network: the one of Modena, Italy. As can be seen in the top-left panel of figure \ref{fig:modena}, real cities are far from being random Erdos-Rényi graphs, since they are mostly planar graphs. We took Modena topology as well as pressure and demand data from the Internet \cite{Modena}, and used EPANET to solve the stationary state of the network. With the resulting fluid velocities in pipes, we computed the delay times in each network pipe using a discretization of 30 seconds, which is smaller than the fastest pipe (41 seconds), and much smaller than the average pipe delay ($\sim 1100$ seconds, more than 30 times the discretization). Then we constructed our abstract graph representing the city and studied many contamination events on random locations. 

\begin{figure}
\includegraphics[width=0.35\linewidth,angle=270]{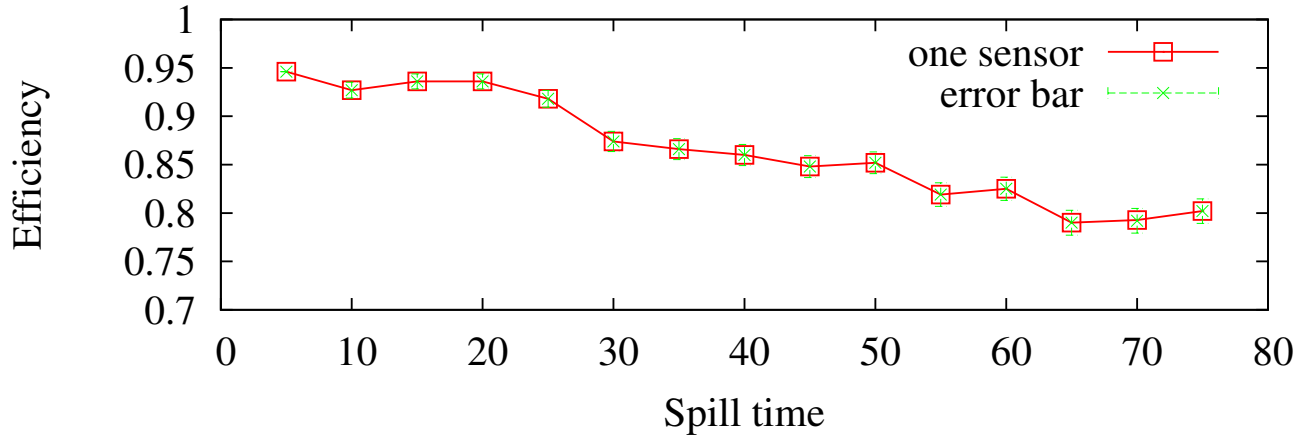}
\caption{\label{fig:eff_vs_spilltime_modena} Efficiency of the method as a function of the duration of the contamination event. The time unit is 30 sec, so the largest spill time considered is 37.5 minutes.}
\end{figure}

We cannot make an efficiency vs size study, since the size of the network is fixed to 268 nodes, but we can study the efficiency with respect to the duration of the source contamination. Notice in figure \ref{fig:eff_vs_spilltime_modena} that for the range from 1 to 75 discrete time intervals, the efficiency  remains above 75\%\ . It seems to be a decreasing function of the contamination time. This is consistent with the fact that the proposed optimization method relies on the assumption that contamination is a rare event. 
This also suggests that a rule of thumb for assuming a good  discretization of time is one that is smaller, but not many times smaller than the typical duration of a contamination event. Of course, the meaning of ``typical'' here will be case specific.

\section{Random Modena-like cities}

We would also like to have statistic results over many different city networks. However, information about the water distribution network of real cities is normally restricted for security reasons. Specially since September 11th terrorists attacks, there has been a growing concern on concealing information that could be used to plan terrorist acts with the highest impact. Unfortunately, such comprehensible precaution affects the testability of studies like this, intended to protect citizens from intentional and non-intentional contamination diffusion.

\begin{figure}
\includegraphics[width=3.5cm,height=3.5cm]{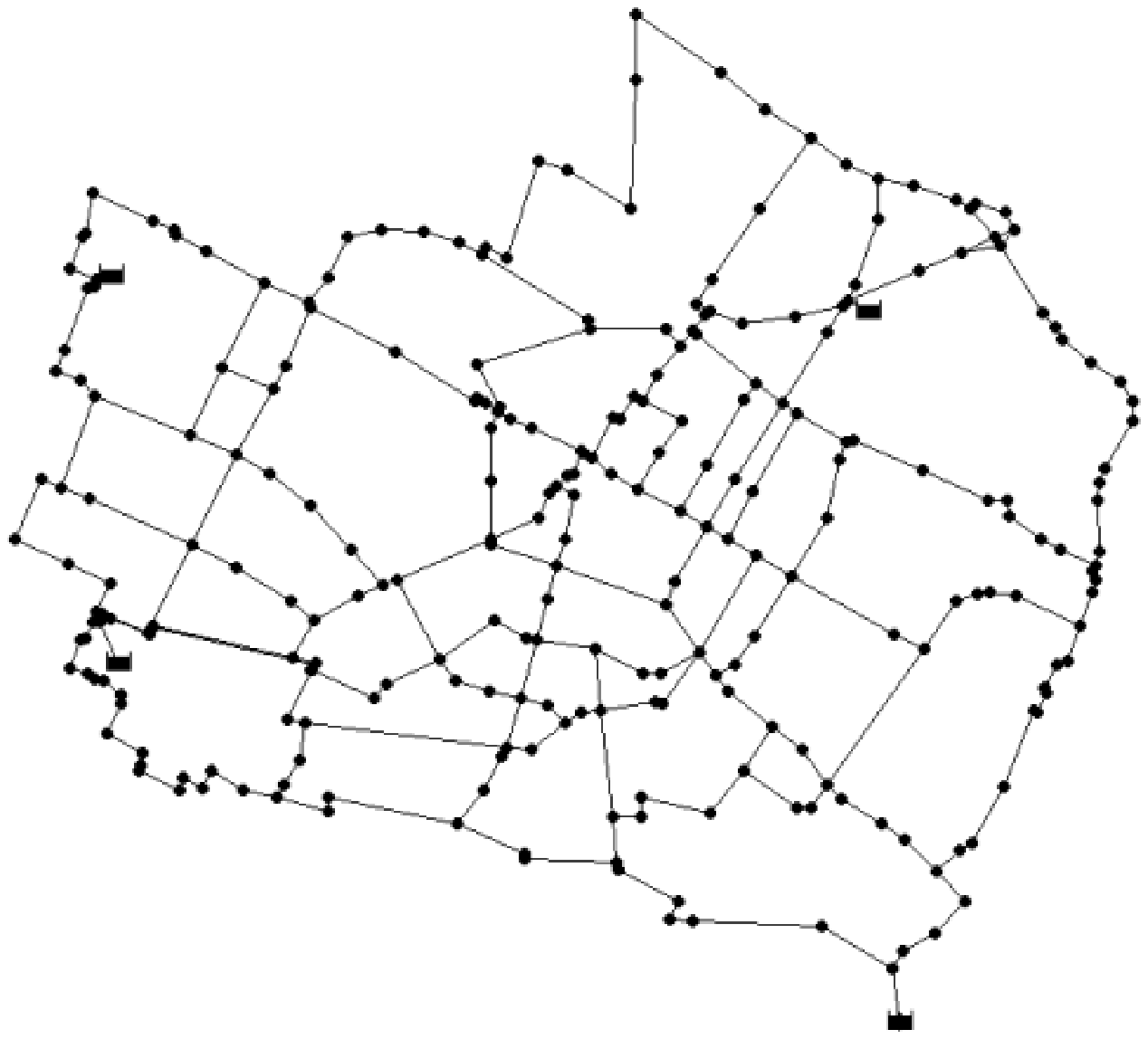}
\includegraphics[width=3.5cm,height=3.5cm]{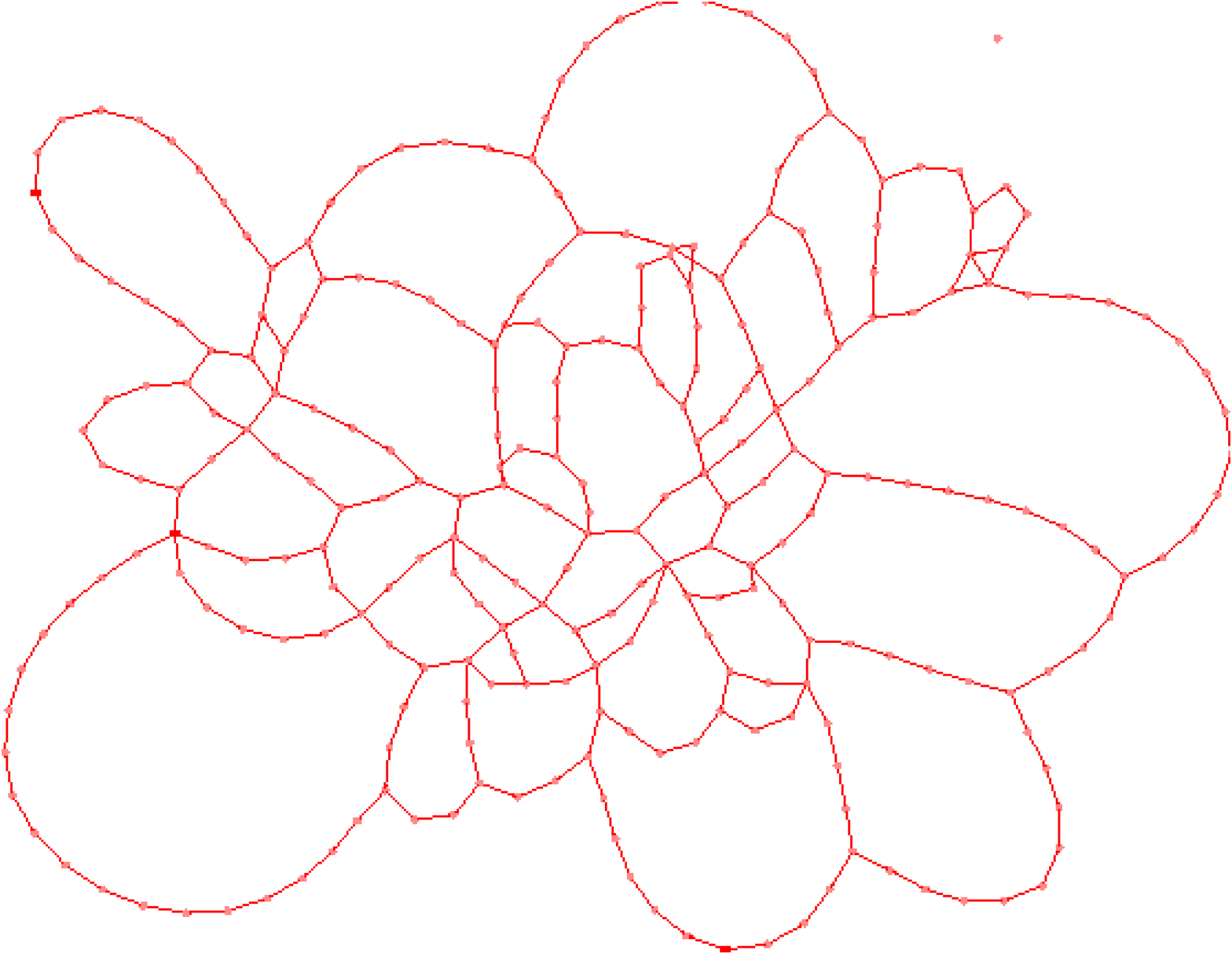}
\includegraphics[width=3.5cm,height=3.5cm]{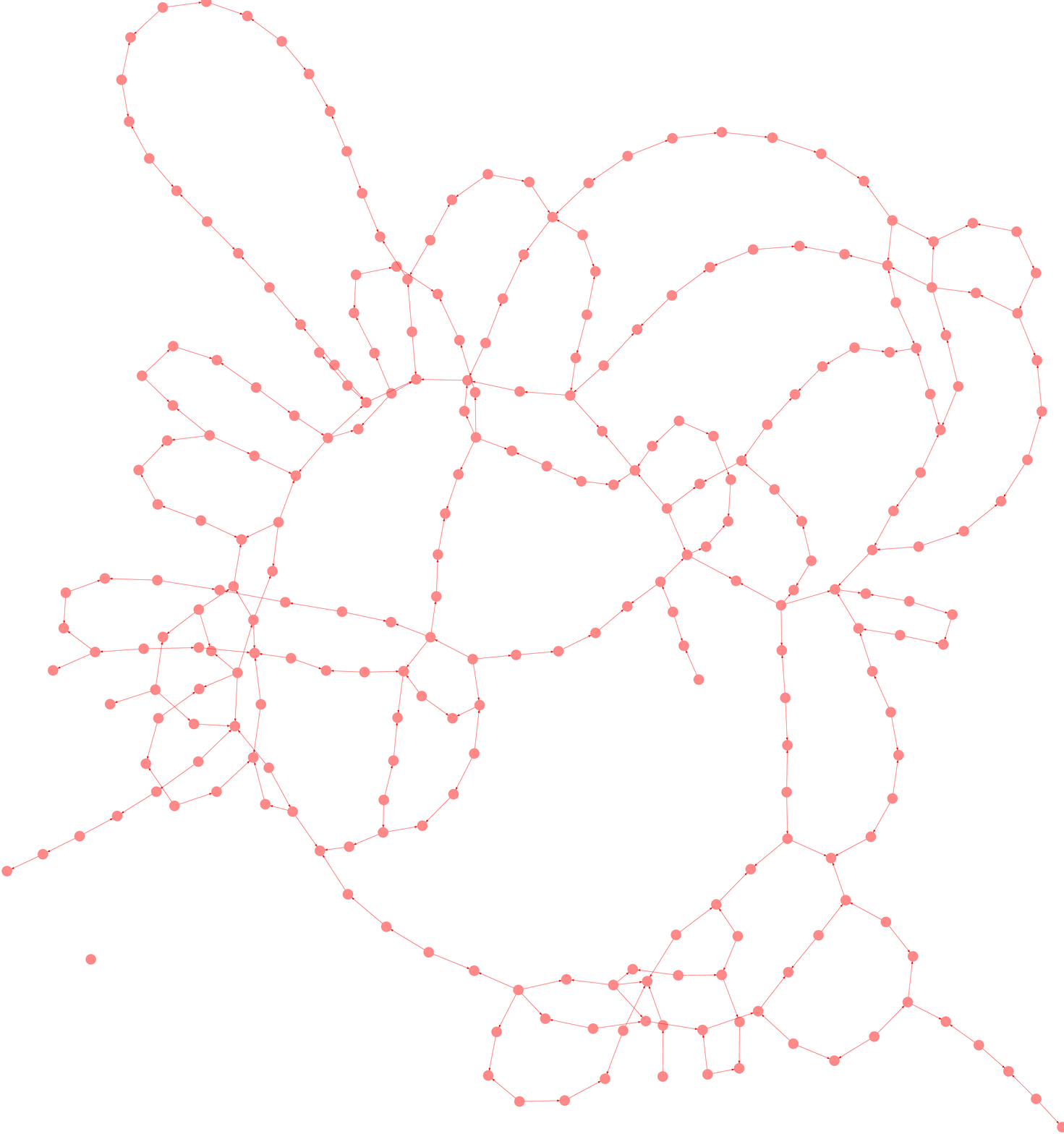}
\includegraphics[width=3.5cm,height=3.5cm]{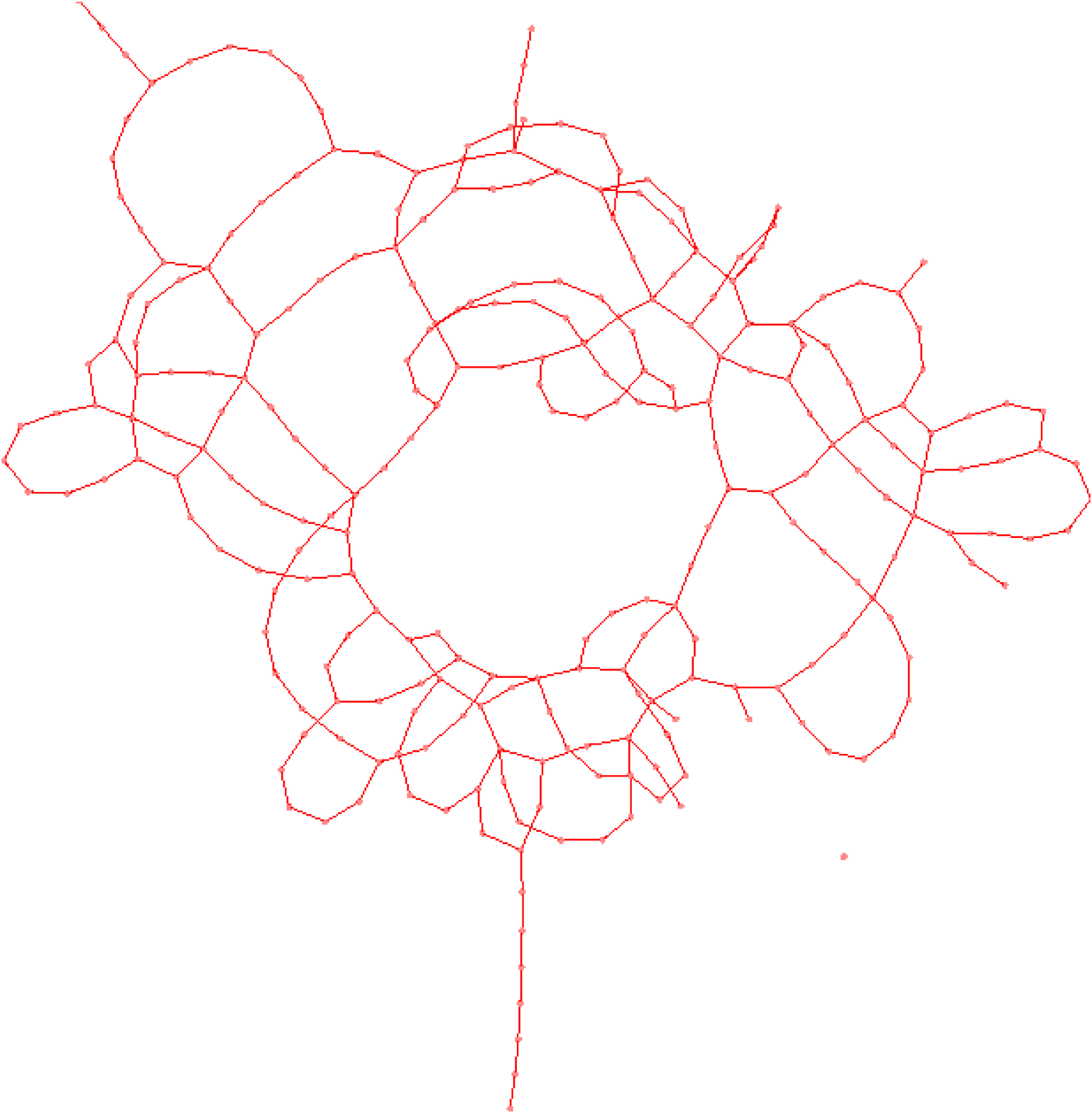}
\caption{\label{fig:modena} Modena and Modena-Like graphs in different views. 1. Graph of Modena city created with EPANET. 2. Graph of Modena city in a different view. 3,4 Graphs of Modena-Like cities.}
\end{figure}

To tackle this we devised a plausible algorithm to create random planar graphs similar to Modena. We start by a square 2D lattice graph of smaller size than Modena. Then we randomize the graph by the following two steps:
\begin{enumerate}
 \item (add nodes) take randomly some edge $(s_i,s_j)$ and replace it by a new node $s_k$ and two new edges $(s_i,s_k)$ and $(s_k,s_j)$. This is repeated until one ends up with a number of nodes similar to that in Modena.
 \item (remove edges) take randomly some edges of the graph and remove them. 
\end{enumerate}
The process is fine-tuned to achieve a similar number of nodes as well as degree distribution as that observed in Modena. The fraction of new nodes introduced to achieve a degree distribution as in Modena is then kept fixed when producing bigger or smaller sizes of networks. Both steps proposed will preserve the planarity of the original 2D lattice. In figure \ref{fig:modena} you can see the similarity between Modena and our random versions of it. The top-left panel represents the real topology of the Modena network, while the top-right one is also Modena, but plotted with a standard graph plotting software. The lower panels are random Modena-like cities produced as explained before.

After creating the artificial Modena-like topology, we randomly assign a rank of pressures to the nodes, and give orientation to the fluid in each pipe accordingly. In the resulting model, we apply a contamination of typical time duration 10 time units to a  random node and evolve it forward in time. Then we try to infer its origin with our optimization method taking the observation at one other random point of the system among those that are touched by the contamination.

With only one sensor, in sizes from 50 to 500 nodes we found that the correct contamination origin was found among the optimal solutions nearly every time (99\%). However, the true solution is not the only one given by the algorithm. In the case with one sensor due to the low connectivity of real networks, and specifically in Modena and Modena-like networks, our method finds many optimal solutions, since almost each time the source and the sensed node are connected by a single path. 

\begin{figure}
\includegraphics[width=0.35\linewidth,angle=270]{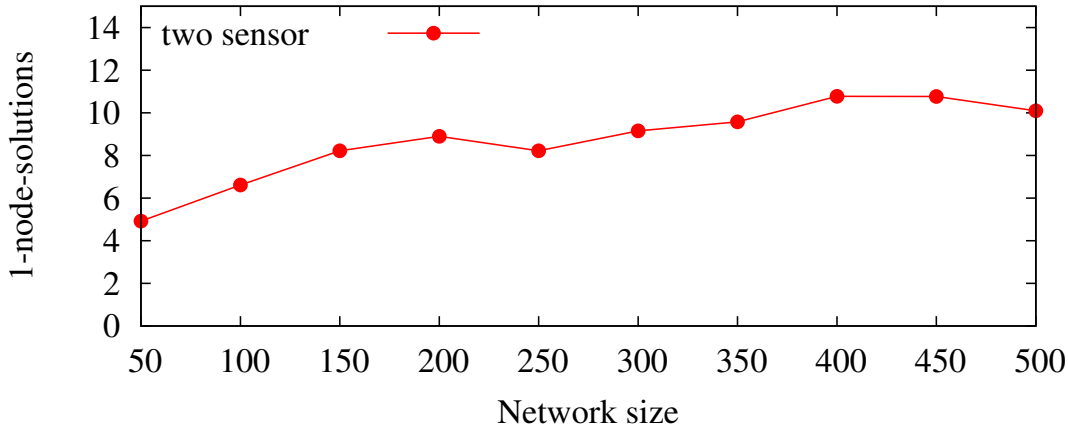}
\caption{\label{fig:network_bi_3} Number of solutions which involve only one node from graph $G$ in Modena-like networks with 2 sensors. Each data point is an average of 1000 different contamination events in different systems.}
\end{figure}

With two sensors, again, the method finds the true contamination pattern almost every time. In figure  \ref{fig:network_bi_3} we show the number of optimal 1-node solutions found, but ``normalizing'' it. As we mentioned, when we have a single line of nodes, the method can not distinguish which among them is the most probable source of contamination. For this reason, we decided to remove from the networks those nodes that have only two neighbors, and replace both pipes by an equivalent larger pipe. The number of 1-node solutions of this reduced version of the network are shown in figure \ref{fig:network_bi_3}.

From the analysis of Modena and Modena like cities, we should emphasize that the method is good at finding the true origin among the solutions, but bad at reducing the number of equivalently good solutions. The multiplicity of optimal solutions is inevitable when the sensor node and the true origin of the contamination are connected by a single path. Let us underline, however, that we are testing our method in random settings, while in real life we expect the positioning of the sensors to have been carefully optimized to reduce the chances of multiple solutions.



 \section{Conclusions}  

In this paper we have presented a method to infer the contamination sources in fluid networks. The method relies on the plausible assumption that contamination events are rare. The inference was formulated in a Bayesian approach and it was transformed into a linear optimization problem with linear restrictions which can be solved using linear programming methods. We emphasize that the Bayesian step is used to formally derive an optimization problem, and not for the treatment of uncertainty in the network parameters (like flow, pipe diameters, etc.) as has been done before. Through the simulation of many random events of contamination in random cities, Modena city and Modena-like cities, it was possible to determine the efficiency of the method (above 70\%), proving the method useful.
 
The method proposed could be readily extended to the case of non-stationary states in the network. If the velocities in the pipes are changing with the demand during the day, for instance, it is not a problem to apply the procedure described to this case. It suffices to create the extended graphs connecting time-space nodes according to the time of the day each node represents.
 
However, the method is too simplistic in many aspects, some of which could be improved. Time discretization is one simplification, but is among those we would rather keep in future treatments. Binary contamination, however, seems a stronger simplification, since a lot of information could be present in the intensity pattern. To keep in the realm of discrete optimization problems, an improvement could be achieved by having more than two discrete contamination levels.
 
Furthermore, we consider that the main drawback of the method for its applicability, is its rigidity. Contamination could pass undetected by observation nodes, specially if highly diluted. In the current setting, any observed non-contaminated time-space node implies the definite elimination of all the upstream nodes as possible origins. We have seen that when using our method with EPANET realistic diffusion, this implies losing the real sources many times. We think that the best improvement over the current method will come from 
applying stochastic methods of inference, like Bayesian networks, to the problem. We are currently moving in this direction.



\bibliography{bibliografia}

\begin{thebibliography}{10}

\bibitem{Ale}
L.~Dall’Asta A. Lage-Castellanos F.~Altarelli, A.~Braunstein and R.~Zecchina.
\newblock Bayesian inference of epidemics on networks via belief propagation.
\newblock {\em Phy.Rev.Lett.}, 112(11):118701, 2014.

\bibitem{luo2013identifying}
W.~P.~Tay W.~Luo and M.~Leng.
\newblock Identifying infection sources and regions in large networks.
\newblock {\em IEEE Trans.Sign.Proc.}, 61(11):2850--2865, 2013.

\bibitem{milling2012identifying}
Sh.~Mannor Ch.~Milling, C.~Caramanis and S.~Shakkottai.
\newblock On identifying the causative network of an epidemic.
\newblock In {\em Communication, Control, and Computing (Allerton), 2012 50th
  Annual Allerton Conference on}, pages 909--914. IEEE, 2012.

\bibitem{spinelli2013source}
B.~M. Spinelli.
\newblock Source detection for large-scale epidemics.
\newblock {\em LCA3, I\&C, EPFL}, 2009.

\bibitem{zhu2016information}
K.~Zhu and L.~Ying.
\newblock Information source detection in the sir model: A sample-path-based
  approach.
\newblock {\em IEEE/ACM Trans.Net.}, 24(1):408--421, 2016.

\bibitem{luo2014identify}
W.~P~Tay W.~Luo and M.~Leng.
\newblock How to identify an infection source with limited observations.
\newblock {\em IEEE Jour.Sel.Top.Sign.Proc.}, 8(4):586--597, 2014.

\bibitem{lokhov2014inferring}
H.~Ohta A.~Y.~Lokhov, M.~M{\'e}zard and L.~Zdeborov{\'a}.
\newblock Inferring the origin of an epidemic with a dynamic message-passing
  algorithm.
\newblock {\em Phy.Rev.E}, 90(1):012801, 2014.

\bibitem{jiang2017identifying}
Sh. Yu Y.~Xiang J.~Jiang, Sh.~Wen and W.~Zhou.
\newblock Identifying propagation sources in networks: State-of-the-art and
  comparative studies.
\newblock {\em IEEE Comm.Sur.\& Tut.}, 19(1):465--481, 2017.

\bibitem{But}
R.~Revelli I.~Butera, F.~Boano and L.~Ridolfi.
\newblock Recovering the release history of a pollutant intrusion into a water
  supply system through a geostatistical approach.
\newblock {\em Jour.Wat.Res.Plan.Manag.}, 139(4):418--425, 2012.

\bibitem{preis2006contamination}
A.~Preis and A.~Ostfeld.
\newblock Contamination source identification in water systems: A hybrid model
  trees--linear programming scheme.
\newblock {\em Jour.Wat.Res.Plan.Manag.}, 132(4):263--273, 2006.

\bibitem{perelman2010bayesian}
A.~L.~Perelman and Ostfeld.
\newblock Bayesian networks for estimating contaminant source and propagation
  in a water distribution system using cluster structure.
\newblock In {\em Water Distribution Systems Analysis 2010}, pages 426--435.
  2010.

\bibitem{cristo2008pollution}
C.~D. Cristo and A.~Leopardi.
\newblock Pollution source identification of accidental contamination in water
  distribution networks.
\newblock {\em Jour.Wat.Res.Plan.Manag.}, 134(2):197--202, 2008.

\bibitem{huang2009data}
J.~J. Huang and E.~A. McBean.
\newblock Data mining to identify contaminant event locations in water
  distribution systems.
\newblock {\em Jour.Wat.Res.Plan.Manag.}, 135(6):466--474, 2009.

\bibitem{hu2015mapreduce}
X.~Yan D.~Zeng Ch.~Hu, J.~Zhao and S.~Guo.
\newblock A mapreduce based parallel niche genetic algorithm for contaminant
  source identification in water distribution network.
\newblock {\em Ad Hoc Net.}, 35:116--126, 2015.

\bibitem{guan}
M.~L. Maslia-W.~M. J.~Guan, M. M.~Aral and Grayman.
\newblock Identification of contaminant sources in water distribution systems
  using simulation--optimization method: case study.
\newblock {\em Jour.Wat.Res.Plan.Manag.}, 132(4):252--262, 2006.

\bibitem{tao2012identification}
T.~Tao, X.~Fu Y-j. Lu, and K~l.~Xin.
\newblock Identification of sources of pollution and contamination in water
  distribution networks based on pattern recognition.
\newblock {\em Jour.Zhej.Univ-Scie.A}, 13(7):559--570, 2012.

\bibitem{wang2011bayesian}
H.~Wang and K.~W. Harrison.
\newblock Bayesian update method for contaminant source characterization in
  water distribution systems.
\newblock {\em Jour.Wat.Res.Plan.Manag.}, 139(1):13--22, 2011.

\bibitem{propato}
M.~M.~Propato, F.~Sarrazy and Tryby.
\newblock Linear algebra and minimum relative entropy to investigate
  contamination events in drinking water systems.
\newblock {\em Jour.Wat.Res.Plan.Manag.}, 136(4):483--492, 2009.

\bibitem{REVERSE}
E.~Salomons and A.~Ostfeld.
\newblock Identification of possible contamination sources using reverse
  hydraulic simulation.
\newblock In {\em 12th Annual International Symposium on Water Distribution
  Systems Analysis, Tucson, Arizona, USA, published on CD}, 2010.

\bibitem{Todini}
E.~Todini and S.~Pilati.
\newblock A gradient algorithm for the analysis of pipe networks.
\newblock In {\em Computer applications in water supply: vol. 1---systems
  analysis and simulation}, pages 1--20. Research Studies Press Ltd., 1988.

\bibitem{Gi09}
D.~Laucelli O.~Giustolisi and A.~F. Colombo.
\newblock Deterministic versus stochastic design of water distribution
  networks.
\newblock {\em Jour.Wat.Res.Plan.Manag.}, 135(2):117--127, 2009.

\bibitem{epanet}
L.~A. Rossman.
\newblock Epanet 2: users manual.
\newblock 2000.

\bibitem{THESIS}
E.~Ortega.
\newblock Inferencia del origen de la contaminación en redes de distribución
  de fluidos.
\newblock Diploma Thesis, Physics Faculty, University of Havana., 2015.

\bibitem{Erd}
P.~Erdos and A.~R\'enyi.
\newblock On random graphs i.
\newblock {\em Publ.Math.Debrecen}, 6:290--297, 1959.

\bibitem{Modena}
University of~Exeter.
\newblock Free {Modena EPANET} model.
\newblock
  http://emps.exeter.ac.uk/media/universityofexeter/emps/research/cws/downloads/data/3-epanet/MOD.inp.
  (2014).

\end{thebibliography}

\end{document}